# Designing Reusable Systems that Can Handle Change
## *Description-Driven Systems : Revisiting Object-Oriented Principles*

Richard McClatchey, Andrew Branson and Jetendr Shamdasani
*Centre for Complex Cooperative Systems, University of the West of England, Bristol, UK*
{Richard.McClatchey, Andrew.Branson, Jetendr.Shamdasani}@cern.ch

Abstract: In the age of the Cloud and so-called 'big data' systems must be increasingly flexible, reconfigurable and adaptable to change in addition to being developed rapidly. As a consequence, designing systems to cater for evolution is becoming critical to their success. To be able to cope with change, systems must have the capability of reuse and the ability to adapt as and when necessary to changes in requirements. Allowing systems to be self-describing is one way to facilitate this. To address the issues of reuse in designing evolvable systems, this paper proposes a so-called *description-driven* approach to systems design. This approach enables new versions of data structures and processes to be created alongside the old, thereby providing a history of changes to the underlying data models and enabling the capture of provenance data. The efficacy of the description-driven approach is exemplified by the CRISTAL project. CRISTAL is based on description-driven design principles; it uses versions of stored descriptions to define various versions of data which can be stored in diverse forms. This paper discusses the need for capturing holistic system description when modelling large-scale distributed systems.

## 1 INTRODUCTION

A crucial factor in the creation of flexible object-based information systems dealing with changing requirements is the suitability of the underlying technology in facilitating the evolution of the system. The importance of clearly defined extensible object oriented models as the basis of rapid systems design has become a pre-requisite to successful systems implementation. Exposing a system's internal architecture opens up its architecture consequently allowing application programs to inspect and alter implicit system aspects. These implicit system elements can serve as the basis for changes and extensions to the system. Making these internal structures explicit allows them to be subject to scrutiny and interrogation.

Related efforts to tackle the problem of coping with design evolution have included, 'active' object models (Yoder & Johnson 2002), the capture and exploitation of so-called mesodata (de Vries & Roddick, 2007), and schema versioning (Roddick, 2009). However, none of these approaches enables the design of an existing system to be changed *dynamically* and for those changes to be reflected in a new running version of that design. We advocate a design and implementation approach that is holistic in nature, viewing the development object-oriented software from a systems standpoint. It is based on the systematic management of the description of essential systems elements covering multiple views of the system under design (including data and process views) using object oriented techniques.

The approach advocated here is termed *description-driven*; it involves identifying and abstracting, at the outset, all the crucial elements (such as business objects, processes, lifecycles, goals, agents and outputs) in the system under design and creating high-level descriptions of these elements which are stored in a model, dynamically modified and managed separately from their instances. In many ways adhering to a description-driven approach means following very closely the original, and these days often neglected or poorly applied, principles of pure object-oriented design especially those of reuse, abstraction, deferred commitment, inheritance and loose coupling.

A Description-Driven System (DDS) makes use of so-called *meta-objects* to store domain-specific system descriptions, which control and manage the life cycles of *meta-object instances*, or domain objects. In a DDS, descriptions are managed independently to allow the descriptions to be specified and to evolve asynchronously from particular instantiations of those descriptions. Separating descriptions from their instantiations allows new versions of items (or item descriptions) to coexist with older versions. This separation is essential in handling the complexity issues facing many computing applications and allows the realization of interoperability, reusability and system evolution since it gives a clear boundary between the application's basic functionalities from its representations and controls.

The next section introduces description-driven systems through an example of their use at the European Centre for Nuclear Research (CERN). The detail of the CRISTAL model is outlined in a later section.

## 2 A DESCRIPTION-DRIVEN SYSTEM IN PRACTICE

Scientists at CERN build and operate complex accelerators and detectors whose construction processes are very data-intensive, highly distributed and ultimately require a computer-based system to manage the production, assembly and calibration of components. In constructing detectors like the Compact Muon Solenoid (CMS, Chatrchyan et al., 2008), scientists require data management systems that can cope with complexity, with system evolution over time and with system scalability.

CMS is a general-purpose experiment that has been constructed from around a million parts and produced and assembled in the past decade by specialized centres distributed worldwide. The construction process was very data-intensive and highly distributed, its production models evolved and required a computer-based system to manage the assembly of detector components. Detector parts of different model versions must be handled over time and coexist with other parts of different model versions. Separating details of model types from the details of parts allowed the model type versions to be specified and managed independently, asynchronously and explicitly from single parts. Moreover, in capturing descriptions separate from their instantiations, system evolution can be catered for while production is underway and provide continuity in the production process and for design changes to be reflected quickly into production.

No commercial products provided the capabilities required by CMS. Consequently, a research project, entitled CRISTAL (Branson et al., 2013) was initiated to facilitate the management of the engineering data collected at each stage of production of CMS. CRISTAL is a distributed product data and workflow management system which makes use of an OO-like database for its repository, a multi-layered architecture for its component abstraction and dynamic object modelling for the design of the objects and components of the system (Estrella, 2001). The DDS approach has been followed to handle the complexity of such a data-intensive system and to provide the flexibility to adapt to the changing scenarios found at CERN which are typical of any research production system. Lack of space prohibits detailed discussion of CRISTAL; a full description can be found in Branson et al., 2013.

The design of the CRISTAL prototype required adaptability over extended timescales for system evolution, interoperability, complexity handling, deferred commitment and for reusability. In adopting a DDS approach the separation of object instances from object description instances was needed. This abstraction resulted in the delivery of a three layer description-driven architecture. Our CRISTAL approach is similar to the familiar model-driven design concepts (OMG, MOF 2004), but differs in that the descriptions and the instances of those descriptions are implemented as objects (Items) and most importantly, they are implemented and maintained using exactly the same internal model. Even though workflow descriptions and instance implementations are different, the manner in which they are stored and are related to each other is the same in CRISTAL. This approach is similar to the distinction between Classes and Objects in the original definition of object oriented principles (Wirfs-Brock et al., 1990). We have followed those fundamental principles in CRISTAL to ensure that we can provide the level of flexibility, maintainability and reusability that object orientation can enable to facilitate system evolution.

## 3 THE CRISTAL MODEL

CRISTAL is an application server that abstracts all of its business objects into workflow-driven, version-controlled 'Items' which are instantiated from descriptions stored in other Items (Figure 1) and are managed on-the-fly for target user communities. Items contain:

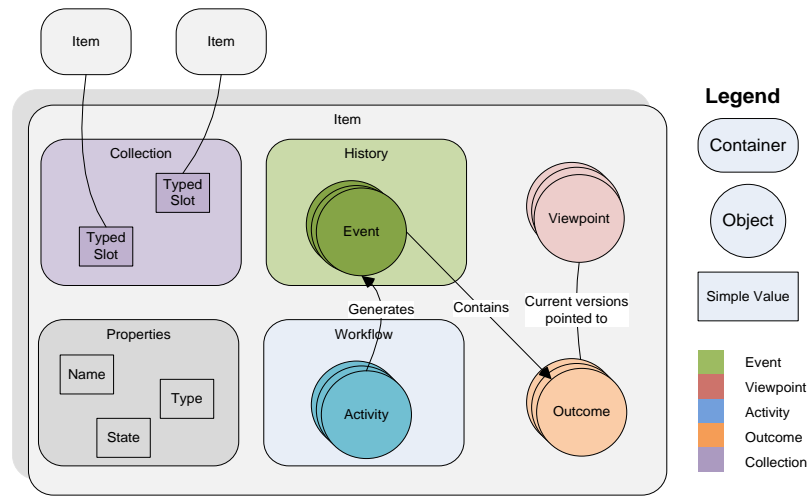

Figure 1. The components of an Item in CRISTAL

- *Workflows*, that comprise of Activities specifying work to be done by Agents (either human users or mechanical/ computational agents via an API), which then generate:
- *Events* that detail each change of state of an Activity. Completion events generate data detailing the work done, known as:
- *Outcomes* which are XML documents from each execution, for which:
- *Viewpoints* refer to particular versions (e.g. the latest version or, in the case of descriptions, a particular version number).
- *Properties* are name/value pairs that name and type items, they also denormalize collected data for more efficient querying, and
- *Collections* that enable items to be linked together.

These Item contents need to be defined when domain systems are modelled in CRISTAL and are, crucially, also modelled using the concept of Items. This is a key difference between DDS and other model driven systems: description items function in exactly the same way as other Items; their workflows consist of activities for managing the data of the description, and also contain an instantiation activity that creates new Items from that data in addition to identifying information for the new Items. The description and its instance share the same implementation, which at any level is capable of being either a model, or an instance, or both. The construction of the specific CRISTAL model for the domain under consideration therefore concentrates on the essential *enterprise objects* of the system that could be needed during its lifetime no matter from which standpoint those objects are accessed. These enterprise objects each have a creation/modification / deletion lifecycle and the CRISTAL model simply keeps track of status changes to the objects (or Items) over those lifecycles. This allows it to orchestrate the execution of Workflows on Items by Agents, log all Events, Outcomes and Viewpoints and thereby capture all associated provenance information associated with the domain system under study.

The basic functionality of CRISTAL is best illustrated with an example: using CRISTAL a user can define product types (such as Newcar spark plug) and products (such as a Newcar spark plug with serial number #123), workflows and activities (to test that the plugs work properly, and mount them into the engine). This allows products that are undergoing workflow activities to be traced and, over time, for new product types (e.g. improved Newcar spark plug) to be defined which are then instantiated as products (e.g. updated Newcar spark plug #124) and traced in parallel to pre-existing ones. The application logic is free to allow or deny the inclusion of older product versions in newer ones (e.g. to use up the old stock of spark plugs). Similarly, versions of the workflow activities can co-exist and be run on these products.

*Item Description* Items hold the templates for new Items, and also dictate their type (see Figure 1). These "Item Descriptions" are also declared as Items (and thus the two can be treated in the same manner), holding the description data as XML outcomes managed through workflow activities. Workflow and Property descriptions are stored as XML serialized objects. Collection Descriptions are themselves Collections, pointing to other Item Descriptions. Outcome Descriptions contain XML

Schema documents which are used to validate submitted outcomes and aid in data collection, for instance to generate data entry forms in a stock GUI for the end users. Also included in the descriptions are Scripts, code invoked by workflows either during a change of Activity state to enact consequences of the execution such as updating a Property or changing a Collection, or to assess conditional splits in the Workflow.

As instances of descriptions can also be descriptions, it is possible to create intermediate description layers that specialize and simplify the architecture of CRISTAL, creating domain specific modelling languages which can flatten the learning curve for domain users and ease adoption. The Agilium system mentioned in section 4 is an example of such a system – it implements BPM as a set of CRISTAL descriptions, and their clients can design and develop applications based on this simpler design language. Writing to the CRISTAL object model is impossible from a client process other than through an activity execution, thus providing full traceability of the system. Ordinary activities only create Events and Outcomes, and modify Viewpoints, so when a script needs to modify some other part of the model it must invoke special 'Predefined Steps' which are activities that contain additional logic for modifying the Item's Properties, Collections or directory entries. These Predefined Steps are hard-coded and do not often change, making their presence in an Item's history reliably interpretable. The aim of this rigidity of write control is to require the design of the lifecycle of each Item type to explicitly define the full behaviour of that Item. We see this as a return to the principles of object modelling that many modern languages and platforms have neglected in the name of "rapid prototyping", whereas a properly designed meta-model should achieve those without sacrificing the principles of object orientation.

At a low-level, the versioning mechanism that gives provenance to the Item instance is the same mechanism that enables concurrent versioning in the descriptions. This means that any communication between different CRISTAL servers can transfer descriptions in exactly the same way as instances. Also dependencies can be declared as easily between abstraction layers as within them. All of these advantages arise because CRISTAL extends the original object orientation concept ideas, to more of its data model than other model-driven systems, in the same way that Java gains similar advantages from implementing classes as Class objects. This is the real benefit of the CRISTAL Item-based design.

A disadvantage to the CRISTAL design is that the definition of 'Object' in the CRISTAL system is an Item which, while adhering to many core concepts of object orientation, does not follow the classic Class/Object model. This is because all Descriptions, and instances of Descriptions, are defined as Items in the CRISTAL model. This was necessary to extend the traceability of the system to its design as well as its operation, and to simplify the styles of objects for developers to master.

Some developers in practice find the abstraction concepts of CRISTAL conceptually difficult to understand. This is due to the large amount of terminology involved in the design of CRISTAL as well as the complexity of its concepts. New personnel faced a steep learning curve before they could usefully contribute to the code-base, though this is not a problem for end-users, as complexity may be hidden in intermediate description layers. However, we feel that Items represent a return to the core values of object orientation, at a time when modern languages are becoming increasingly profligate in their implementation of them in the name of efficiency, thereby sacrificing many of the benefits that object orientation can offer.

Object-orientation encourages the developer to think about the entities involved in the system and the operations required to provide the system's functionality, along with their context in the data model, which together provide the methods of identified data objects, resulting in an object model. In recent years, newer programming languages have tended to focus on object orientation as a means of API specification, increasing the richness of library specification and maximizing code reuse, but do little to encourage proper object oriented design amongst developers. Unfortunately, with the increasing popularity of test oriented development methodologies, developers are encouraged to hack away in a deliver-early-and-often way from which a well-thought out object model rarely emerges.

In contrast with CRISTAL the object model must be designed as a set of Items with lifecycles. While other non-Item oriented software components are possible, they cannot store state in the system without interacting with Item activities, and therefore are encapsulated as Agent implementations, and considered external to the Item model, with a strictly designed outcome specification stating what they must provide to the system to have successfully completed their function. The activities of an Item's lifecycle are roughly analogous to object oriented methods, since they define a single action performed on that Item.

However, it is much harder for an Item's lifecycle design to grow out of control with many unused methods since the lifecycle is defined as a workflow; the activity set must always form a valid graph of activities from the creation of the Item to its completion. This clarity of design through implementation constraints is a return to the intentions of the early object oriented languages such as Smalltalk (Goldberg et al, 1983), and the initial restrictions of Java, which discouraged the developer from using mechanisms that could result in messy, overcomplicated, unmaintainable code, and steer them towards a core object oriented design with the system logic intuitively partitioned and distributed in a manageable way.

The CMS Electromagnetic Calorimeter (ECal) was constructed from tens of thousands of similar parts, monocrystals of lead tungstate to be exact, all needing characterizing and assembling in an optimal configuration based on sets of detailed measurements. These characterizations are used in the final operation of the ECal to determine physical measurements in the CMS detector. Every component part was registered as an Item in the CRISTAL database, each with its barcode as an identifier. Each part had a type, which functioned as the Item Description, and was linked to the Workflow definition that each instance would follow in order to collect its data and mount sub-parts (Estrella, 2003). The part types also contained subtype data as Properties and Collection Definitions to make sure that parts were assembled in assigned positions in ECal. All collected assembly data were stored as Outcomes attached to Events, and therefore, the entire history of every interaction with the application was recorded. The result was a set of *Items* representing the top level components of the detector which contained five levels of substructure, all with their full production history and with all collected and calculated production data attached in the correct context.

## 4 AN EVALUATION OF THE APPROACH USED IN CRISTAL

Each ECal crystal generated between 2-3Mbytes of information which was mainly gathered in an automated data acquisition system which characterised the crystals in batches over a period of 8-10 hours for each batch of 30 crystals. The whole data acquisition process took around five years to complete following an initial testing period which itself took several months. It was the responsibility of one CRISTAL application maintainer to ensure as smooth operation as possible of the data acquisition and to provide round-the-clock accessibility to the CRISTAL database and to maintain the descriptions handled by CRISTAL.

During the six years of near-continuous operation, the descriptions went from beta to production then through years of (relatively few) alterations of the domain logic which necessitated very little change in the actual server software, illustrating the flexibility of the CRISTAL approach (see Table 1). These alterations were minor and included updates to descriptions of processes and data sources which were handled by version management capability of CRISTAL. The server software only needed to be upgraded seven times, and of those seven, only one was a required update that needed to be made available to all users and servers. This was necessary because some data formats originally designed proved not to be as scalable as required; therefore a client update was required to read the new structures.

The application logic that needs to be executed during the workflow will have its functionality conveniently broken down along with the activities. It is then simple to import these definitions into the system where it can be immediately tested for feedback to the users. Improvements can thereby be quickly performed online, often by modifying the workflow of one test item, which then serves as a template for the type definitions. Items subject to the improvements can co-exist with items generated earlier and prior to the improvement being made and both are accessed in a consistent, reusable and seamless manner. All this can be done without recompiling a single line of code or restarting the application server, providing significant savings in time and enables the users to work in an iterative and reactive manner that suits their research. This shows the flexibility of using a DDS approach.

In our experience, the process of factoring the lifecycle and dataset of the new item type into activities and outcomes helps to formalize the desired functionality in the user's mind; it becomes more concrete - avoiding much of the vague and often inconclusive discussion that can accompany user requirements capture. Because it evolved from a production workflow specification driven by user requirements, rather than a desire simply to create a 'workflow programming language', CRISTAL's style of workflow correlates more closely to the users' concept of the activities required in the domain item's lifecycle. The degree of granularity can be chosen to ensure that the user feels it provides sufficient control, with the remaining potential subtasks rolled up into a single script. This is one important aspect of the novel approach adopted during CRISTAL development that has proven of benefit to its end-user community. In practice this has been verified over a period of more than 10 years use of CRISTAL at CERN and by its exploitation as the Agilium product (Agilium, 2008) across many different application domains in industry (see discussion in the later conclusions section).

After its development at CERN, many different features have been added to CRISTAL. One example of this is to facilitate the extensibility of CRISTAL by having a pluggable architecture based on *modules*. Originally, CRISTAL could support only one domain application per instance, but using CRISTAL modules, many different groupings of functionalities can be loaded in the same instance. Modules may declare themselves dependent on each other when they rely on or extend functionality from other modules, thereby, allowing extensibility of the system. The module itself is abstracted as an Item in each system into which it is loaded, and so is versioned and traced. This mechanism makes it possible to have description-driven libraries. This extensibility is arguably the main contribution since the CRISTAL developments carried out at CERN. It has provided us with a means to have a pluggable architecture and is closer to the definition of reuse in the original OO model. Certainly the main lesson learnt from the CRISTAL project in coping with change was to develop a data model that had the capacity to cover multiple types of data (be they products or activities, atomic or composite in nature) and at the same time was elegant in its simplicity. To do this a disciplined and rigorously applied object-oriented approach to data modelling was required: designers needed to think in a way that would ultimately facilitate system flexibility, would enable rapid change and would ease the burden of maintenance from the outset of the design process.

The approach that was followed in designing CRISTAL was to concentrate on the essential enterprise objects and descriptions that could be needed during the lifetime of the system no matter from which standpoint that data is accessed.

Thus the system was allowed to be open in design and flexible in nature and the elegance of its design was not compromised by being viewed from one or several application-led standpoints (such as Business Process Management (BPM Weske, 2007), Workflow Management Systems (WfMS Georgakopoulos, 1995) or many others. Rather we enabled the traceability of the essential enterprise objects over the lifetime of the system as the primary goal of the system and left the application-specific views to be defined as and when they became required. The ability of description-driven systems to both cope with change and to provide traceability of such changes (i.e. the 'provenance' of the change) we see as one of the main contributions of the CRISTAL approach to building flexible and maintainable systems and we believe this makes a significant contribution to how enterprise systems can be implemented. For more detail, consult our previous paper (McClatchey, 2013) which discusses this in a practical application. Recently a start-up company called Technoledge has been established to develop applications of CRISTAL.

These design skills were not simple; designers needed to be able to think conceptually, abstracting the characteristics of everyday objects into 'items' with associated metadata and to be able to represent that complexity in a concrete data model. Great benefits in terms of maintainability and flexibility resulted from being able to treat many different system objects in a single standardised manner. Savings over the lifetime of the ECAL project at CERN are estimated at several man years of effort. The importance of instantiation and description in formulating a generic CRISTAL data model cannot be overemphasised. We propose that the description-

| Table 1 - Statistics of CRISTAL operation at CERN CMS ECal **Global ECal CRISTAL Statistics** | |
|---|---|
| Total number of centres (servers) | 9 (6 at CERN, 1 in Taiwan, 2 in Greece) |
| Runtime | August 2003 – August 2009 (6 years) |
| Total data size (at CERN) | 210GB |
| Total number of Items in one ECAL | 450,000 |
| Minor version upgrades (required client update) | 1 |
| Total number of kernel builds | 22 |
| Kernel builds requiring server software upgrade | 7 |

driven design approach that emerges from this study is a genuinely new approach to designing for change.

Great importance was placed on the involvement of users at all stages of the development of CRISTAL, following many of the principles of participatory design (Kensing and Blomberg, 1998). We regard this as one of the prime reasons for the eventual success of the project. The research nature of the environment in which CRISTAL was formulated and developed led to both advantages and disadvantages. Although initially it was hoped that high-end expert users would be able to develop workflows themselves, in practice this was not possible. Instead the users collaborated closely with the designers from the outset of the project to establish a much clearer idea of the implications of their requirements, and with a full understanding of the functionality that their workflow must provide. This could then be implemented with verifiable accuracy to what the user originally specified.

Essentially this approach led to a very simple way of representing new requirements and absorbing them rapidly into the evolving data model, as and when they emerged. On the negative side users necessarily did not always know at the outset what their final requirements would be for data and process management, leading to disruptive changes in design decisions and an evolutionary approach to prototyping. On the positive side, the users were not locked into a 'static' product: the CRISTAL model evolved to cater for their requirements and was made responsive to their needs.

Control of evolving user requirements was a particularly challenging problem. New requirements needed to be addressed at the application level which, as a consequence, induced requirements at the domain implementation level which in turn passes its own requirements down to the kernel level. The result of this was that there could be a considerable number of potential feature configurations of the CRISTAL kernel needed to meet all possible requirements from the user. Since CRISTAL was originally conceived as an object-based system and an object-oriented approach was adopted in its design, an attempt was made to follow as far as was practically possible best software engineering practice in implementing features associated with object oriented models in order to ensure reuse and extensibility. Whenever a new design modification was needed, the approach taken was always to implement as open and flexible a solution as the design allowed in order not to constrain future extensions.

In practice, however, this quickly led to spiralling complexity and to a risk of compromising the system development process. To address this situation the approach that we adopted was to make the implementation of new requirements as intuitive as possible with as simple functionality as necessary to cope with the requirements, thereby preserving the elegance of the original (description-driven) design. This led to a closely connected set of system functionalities which was easy to maintain and to dynamically extend when required. In addition this much simpler system has the virtue of being a lot easier for users, developers and administrators new to the system to pick up and start working with.

Further evidence of the benefits accruing from use of CRISTAL comes from its commercialization as the Agilium product. Since 2004 an early version of the CRISTAL Kernel has been exploited by the M1i company (based in Annecy, France) for the purpose of supporting BPM and the integration and co-operation of multiple business processes especially in business-to-business applications. M1i have taken CRISTAL and added applications for BPM that benefit from the description-driven aspects of CRISTAL, i.e. its flexibility, reusability, complexity handling and system evolution management. Their product addresses the harmonization of business processes by the use of a CRISTAL database so that multiple potentially heterogeneous processes can be integrated and have their workflows tracked in the database. Agilium also integrates the management of data coming from different sources and unites BPM with Business Activity Management (BAM) (Kolar, 2009) and Enterprise Application Integration through the capture and management of their designs in the CRISTAL system. Using the facilities for description and dynamic modification in CRISTAL, Agilium is able to provide modifiable and reconfigurable business workflows. Details of Agilium can be found at (Agilium, 2008).

## 5 CONCLUSIONS

The study described in this paper has demonstrated the benefits of a self-describing description-driven design approach to both designer and to users in practice. It has shown that describing a proposed system explicitly and openly from the outset of the project enables the developer to change aspects of it responsively as users' requirements evolve. This enables seamless transition from version to version with (virtually) uninterrupted system availability and facilitates full traceability throughout the system lifecycle.

Following the principles of object-oriented design the approach encourages reuse of code, configuration data and scripts/methods. Indeed, the description-driven design approach takes this one step further and provides reuse of meta-data, design patterns and maintenance of items and activities (and their descriptions). Practically this results in a higher level of control over design evolution and simpler implementation of system improvements and easier maintenance cycles. Many system elements have gained in conceptual simplicity and consequent ease of management thanks to loose typing and the adoption of a unified approach to their online manipulation: activities/scripts and their methods; member types and instances; properties and primitives; items and collections; and outcome schemas and views. One logical consequence of providing such a unified design and simplicity of management is that the CRISTAL software can be used for a wide spectrum of application domains.

Future work is being to model domain semantics e.g. the specifics of a particular application domain e.g. healthcare, public sector, finance, and aerospace. This will essentially transform CRISTAL into a self-describing model execution engine, making it possible to build applications directly on top of the design, without code generation. The design will be the framework for all of the application logic – without the risks of misalignment and subsequent loss that code generation can bring – and for CRISTAL to be configured as needed to support the application logic whatever it may be. What this means is that the CRISTAL kernel will be able to capture information about the application area in which a particular instance is being used. This will allow usage patterns to be described and captured, roles and agents to be defined on a per-application basis, and rules and outcomes specific to particular user domains to be managed. This will enable multiple instances of CRISTAL to discover the semantics required to inter-operate and to exchange data. Research into the further extension and uses of CRISTAL continues. There are plans to enrich its kernel (the data model) to model not only data and processes (products and activities as items) but also to model agents and users of the system (whether human or computational). It is planned to investigate how the semantics of CRISTAL items and agents could be captured in terms of ontologies and thus mapped onto or merged with existing ontologies for the benefit of new domain models. The emerging technology of cloud computing and its application in complex domains, such as medicine and healthcare, provide further interesting challenges.

## ACKNOWLEDGEMENTS

The authors wish to highlight the support of their home institute across all of the projects that led to this paper.